\documentclass[a4paper]{article}

\usepackage{INTERSPEECH2022}
\usepackage{colortbl}

\title{InQSS: a speech intelligibility and quality assessment model using a multi-task learning network}

\name{Yu-Wen Chen, Yu Tsao}
\address{Research Center for Information Technology Innovation, Academia Sinica, Taiwan}
  
\email{}

\begin{document}

\maketitle
\begin{abstract}
Speech intelligibility and quality assessment models are essential tools for researchers to evaluate and improve speech processing models. However, only a few studies have investigated multi-task models for intelligibility and quality assessment due to the limitations of available data. In this study, we released TMHINT-QI, the first Chinese speech dataset that records the quality and intelligibility scores of clean, noisy, and enhanced utterances. Then, we propose InQSS, a non-intrusive multi-task learning framework for intelligibility and quality assessment. We evaluated the InQSS on both the training-from-scratch and the pretrained models. The experimental results confirm the effectiveness of the InQSS framework. In addition, the resulting model can predict not only the intelligibility scores but also the quality scores of a speech signal.

\end{abstract}


\section{Introduction}
\label{introduction}
The speech intelligibility and quality assessment models are important because they provide an efficient way to evaluate the speech processing models, such as the speech enhancement (SE), automatic speech recognition (ASR), and voice conversion models. Various methods have been introduced to measure the speech intelligibility \cite{taal2011algorithm, jorgensen2013multi, van2017instrumental} and quality \cite{rix2001perceptual, huber2006pemo, visqol43990}. However, these methods are intrusive measurements that require knowing the corresponding clean speech signals, which are often unavailable in real-world applications. In addition, the results might not correlate well with the listening test results \cite{reddy2019scalable, li2020noise, reddy2021dnsmos}. 

To address this problem, recent studies have used listening test datasets to build non-intrusive neural network models that can predict human perception. For example, \cite{mosnet, leng2021mbnet, choi2021neural, tseng2021utilizing} proposed quality assessment models trained on VCC challenge dataset and \cite{reddy2021dnsmos} proposed DNSMOS trained on the DNS challenge dataset \cite{reddy2020interspeech}. For intelligibility assessment, the models in \cite{ spille2018predicting} and \cite{arai2019predicting} were trained using the dataset presented in \cite{schubotz2016monaural} and a self-collected Japanese corpus, respectively. In addition, \cite{pranay2021noresqa} used previous quality assessment models as references, and \cite{zhang2021end} proposed a multi-task model trained on subjective quality scores and objective scores. However, these models can only perform subjective intelligibility or quality assessment but not both. Furthermore, previous datasets are mostly in English, so datasets in other languages are required. 

The contributions of this study are as follows:
\begin{itemize}
\item We released TMHINT-QI\footnote{Download TMHINT-QI: https://github.com/yuwchen/InQSS}, a Chinese speech dataset with subjective quality and intelligibility scores. The dataset includes clean, noisy, and enhanced noisy utterances processed using SE models. To the best of our knowledge, this is the first Chinese dataset that records the quality and intelligibility scores of enhanced speech signals.

\item We propose the InQSS, a multi-task learning framework using training-from-scratch and pretrained self-supervised learning (SSL) models. The experimental results confirmed the effectiveness of incorporating the training of quality and intelligibility predictions in the same model. In addition, the resulting models are the first multi-task intelligibility and quality assessment models trained on the subjective quality and intelligibility scores. 

\item We tested and demonstrated the effectiveness of using scattering coefficients \cite{mallat2012group}, which are helpful for several signal processing tasks \cite{anden2014deep, zeghidour2016deep, ghezaiel2021hybrid} but have not yet been tested in a speech assessment model.

\end{itemize}

\section{TMHINT-QI dataset}
\label{sec:dataset}
The TMHINT-QI dataset is a Chinese speech dataset with subjective quality and intelligibility scores. TMHINT-QI includes clean utterances recorded in a quiet environment, artificially contaminated noisy utterances, and enhanced noisy utterances processed by the SE models. The goal of releasing TMHINT-QI is to facilitate research on speech quality and intelligibility assessment models, which can later be used to assess the performance of speech processing, thereby guiding researchers to develop models that can generate results with better human perception.

\subsection{Data preparation}

The TMHINT-QI dataset used the TMHINT sentences \cite{huang2005development} and was recorded in a 16-bit format at a 16-kHz sampling rate. Each utterance contained 10 Chinese characters of approximately 3 s in duration and was recorded in a quiet room. The recorded utterances were divided into two parts.

The first part consisted of 3 female and 3 male speakers, each reading 200 sentences with a total of 1200 clean utterances. To form noisy utterances, each clean utterance was contaminated with five randomly sampled noise types from a 100-noise type dataset \cite{hu2004100} at 8 different SNR levels ($\pm$1 dB, $\pm$ 4 dB, $\pm$ 7 dB, and $\pm$ 10 dB). These clean-noisy paired utterances were then farther used as training data for the three neural-network-based SE models including FCN \cite{fu2017raw}, DDAE \cite{lu2013speech}, and transformer-based SE \cite{kim2020t} (denoted as Trans). 

The second part of the recorded TMHINT utterances contained one female and one male speaker. Each speaker recorded 115 sentences, a total of 230 clean utterances were recorded. The second part of the utterances was used for the formal listening test. The recorded clean utterances were contaminated with four types of noise (babble, street, pink, and white) at four SNR levels (-2, 0, 2, and 5) to form the noisy speech. These noisy speeches were then processed using five SE models (KLT \cite{rezayee2001adaptive}, MMSE \cite{ephraim1985speech}, FCN, DDAE, and Trans). Finally, the clean, noisy, and enhanced utterances were combined to form the listening test utterance pool.

\subsection{Listening test}

In the listening test, the participants scored the utterances based on quality and intelligibility. The quality score is on a scale of 1-5, where 1 represents the lowest perceived quality and 5 indicates the highest perceived quality. The intelligibility score is the number of characters a participant can recognize from an utterance. Because each TMHINT sentence contains 10 characters, the intelligibility score is within the range of 0-10, where the score represents the number of characters a participant can recognize in the sentence. 


The listening test was divided into pretest and formal test. In the pretest, the participants had to listen to and score five clean utterances from the SE models' training data. We informed the participants that these clean utterances had a quality score of 5, so the upper bound of quality scores might be less affected by the headphones they used. In addition, participants' ability to recognize all characters of clean utterances provided an indicator for evaluating their understanding of Chinese and hearing status.

During the formal test, the participants listened to and scored 103 utterances chosen from the listening test utterance pool. The samples in the formal test started with two randomly chosen files, and the other 101 files included 5 clean utterances and 96 noisy or enhanced utterances. To avoid the problem of a data imbalance in later studies, the number of testing files was designed to be 96 such that each participant would score the same number of files under different SNRs, methods, or noise types. For example, because there are six processing methods (five SE methods and one without processing), a participant listened to 16 (96/6) samples for each method. 


In total, TMHINT-QI contains 24,408 samples with 14,919 unique utterances (including example files). These data were collected from 226 participants (94 males and 132 females) between 20 and 50 years of age.

\subsection{Characteristics of TMHINT-QI}

\subsubsection{Comparison between objective assessment metrics and the listening test}

\begin{figure*}
\centerline{\includegraphics[scale=0.8]{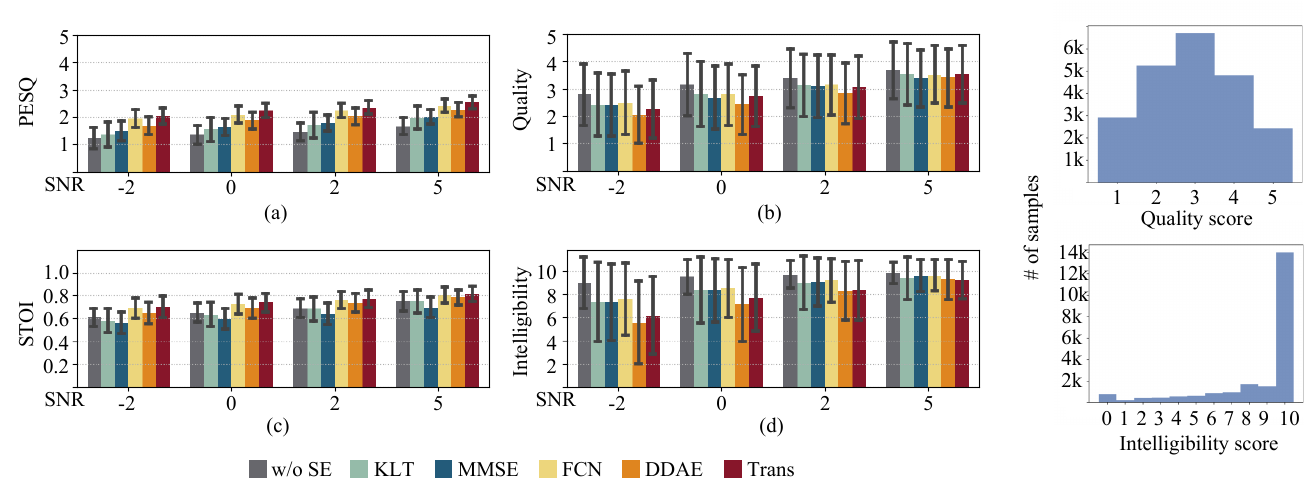}}
\vspace{-0.5cm}
\caption{Comparison between objective assessment metrics and the listening test (left), and histograms of the quality scores and the intelligibility scores (right).}
\label{dataset_info}
\end{figure*}

We compared two objective assessment metrics (PESQ \cite{rix2001perceptual} for quality and STOI \cite{taal2011algorithm} for intelligibility) with the listening test results. Figure~\ref{dataset_info} (left) shows the average PESQ, STOI, quality scores, and intelligibility scores of the SE methods under different SNRs. For PESQ and STOI, the results in (a) and (c) show that noisy utterances without SE (w/o SE) achieved the worst performance, and neural network-based SE methods (FCN, DDAE, and Trans) outperformed the traditional SE methods (KLT and MMSE). These results are also in line with previous studies and our expectations. However, unlike the PESQ and STOI results, the listening results in (b) and (d) show that noisy utterances without applying SE achieved the highest scores in both quality and intelligibility. 

The results of our listening test align with the finding of previous studies \cite{reddy2019scalable, reddy2021dnsmos, li2020noise} that the objective assessment metrics might not correlate well with human perception. Note that in the listening test results, the higher the SNRs were, the higher the quality and intelligibility scores the utterances received. This result matches the intuition; thus, we believe the collected scores can reflect general perceptions to some extent.


\subsubsection{Characteristics of quality and intelligibility scores}

Figure~\ref{dataset_info} (right) shows the histograms of the quality and the intelligibility scores, which exclude the scores of clean utterances. The quality scores were distributed close to a Gaussian with a mean of approximately 3. The distribution of intelligibility scores had a strong left skewness and a peak at the maximum. This result indicates that the participants can recognize whole sentences in most samples. In addition, we found similar distributions of the quality and intelligibility scores in the VCC 2018 \cite{mosnet} and ADFD datasets \cite{andersen2018nonintrusive}, respectively. 

\subsubsection{Correlation between quality and intelligibility scores}

\begin{figure}[htbp!]
\centerline{\includegraphics[scale=0.85]{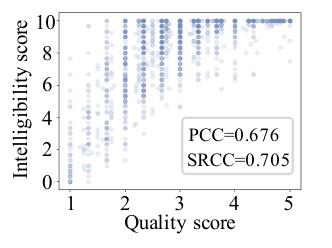}}
\vspace{-0.4cm}
\caption{Scatter plot of intelligibility scores versus quality scores. Each dot represents a sample point. The darker the color, the more samples that are gathered.}
\label{corr}
\end{figure}

Figure~\ref{corr} shows the correlation between intelligibility and quality scores of testing data in \ref{sec:experiment_settings}. The high correlation between the intelligibility and quality scores motivates the study of using a multi-task learning network to enhance the performance of a single task.


\section{Proposed InQSS}
\label{sec:proposed}

\subsection{Related works}
\subsubsection{Scattering transform}
The scattering transform was proposed in \cite{mallat2012group} for building robust, time-shift-invariant, and informative signal features. Previous studies have shown that such features can be successfully applied in several signal processing tasks \cite{anden2014deep, zeghidour2016deep, ghezaiel2021hybrid}. However, scattering coefficients have not yet been tested in speech assessment. 
 
\subsubsection{MOSNet}
MOSNet \cite{mosnet} is a non-intrusive quality assessment model based on convolutional neural network (CNN) and bidirectional long short-term memory (BLSTM). In addition, MOSNet uses a combination of utterance- and frame-level losses such that the prediction is more correlated with human ratings \cite{fu2018quality}. 

\subsubsection{SSL}

SSL aims to generate a general representation of the input signal \cite{liu2022audio}. Previous studies have demonstrated the strong performance of using SSL features as additional inputs \cite {tseng2021utilizing} or fine-tuning SSL models for various tasks \cite{siriwardhana2020jointly, cooper2021generalization}.

\subsection{InQSS framework}

We propose InQSS, a multi-task learning framework for Intelligibility and Quality aSSessment. We tested the framework on three model structures, including a model trained from scratch (InQSS-MOSNet), a model fine-tuned from a pretrained SSL (InQSS-SSL), and an ensemble model (InQSS-MOSSSL).

\begin{figure}[htbp!]
\centerline{\includegraphics[scale=0.85]{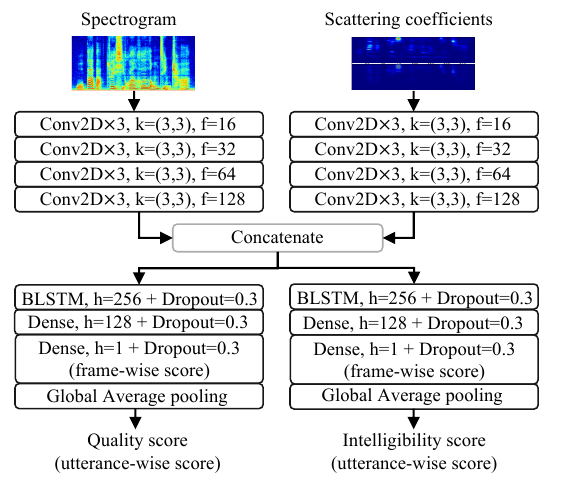}}
\caption{Model structure of InQSS-MOSNet. In convolution layers, \emph{k} and \emph{f} denote the kernel size and the number of filters, respectively. In the BLSTMs and dense layers, \emph{h} denotes the hidden size. Note that a dense layer is a time-distributed dense layer, and the ReLU activation function following every convolution and dense layer is not shown in the figure.}
\label{mosnet_structure}
\end{figure}

The InQSS-MOSNet is based on the MOSNet. We use the same utterance- and frame-wise losses and a similar CNN-BLSTM structure. We improve the MOSNet by adding a multi-task structure and incorporating the scattering coefficients as additional input features. The input spectrogram and scattering coefficients, which are a concatenation of first- and second-order scattering coefficients, first pass through two separated CNNs and are then concatenated as the input of the BLSTMs. Finally, the outputs of the BLSTMs go through the dense layers and give the predicted scores. The InQSS-MOSNet is trained with the L2 loss of intelligibility and quality scores. The model structure of InQSS-MOSNet is shown in Figure~\ref{mosnet_structure}.

\begin{figure}[htbp!]
\centerline{\includegraphics[scale=0.85]{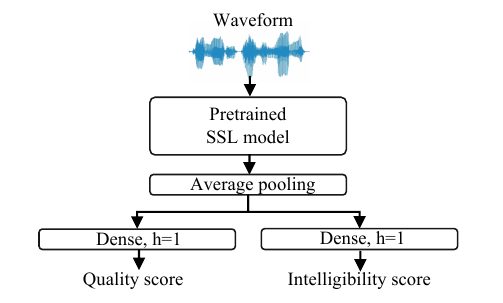}}
\caption{Model structure of InQSS-SSL. In this study, the SSL model we used is \emph{w2v\_small} in the Fairseq toolkit \cite{ott2019fairseq}.}
\label{ssl_structure}
\end{figure}

For the InQSS-SSL, we fine-tune a pretrained SSL model by average-pooling the model’s output embeddings and adding a dense output layer for intelligibility and quality prediction. Then, the InQSS-SSL is trained with the L1 loss of intelligibility and quality scores. The model structure of InQSS-SSL is shown in Figure~\ref{ssl_structure}.
In InQSS-MOSNet and InQSS-SSL, both the prediction paths are trained simultaneously. For the InQSS-MOSSSL, we first finish the training of the InQSS-MOSNet and InQSS-SSL separately. Then, we average the predictions as final results. 

\section{Experiments}
\label{sec:experiment}
\subsection{Experimental settings}
\label{sec:experiment_settings}

In this study, we used the TMHINI-QI dataset to evaluate the proposed assessment models. To reduce the effect of the bias of the participants on the assessment, we collected utterances with at least three sampled scores as testing data. Then, we used the average scores as target scores. In total, the testing set contained 1978 unique utterances. The remaining 17,448 samples in TMHINT-QI were used for training. Note that the example files in TMHINT-QI were excluded from the testing data and only used for training. During training, we randomly sampled 10\% of the data for validation and used these validation data as a reference for early stopping and model selection. The model performances in the following are the average performance of four models trained with different training and validation splits.

The spectrogram was calculated using the short-time Fourier transform with a window length of 512 and a hop length of 256 samples. We used the Kymatio \cite{andreux2020kymatio} toolkit to conduct scattering transform. The scattering coefficients were calculated by setting the resolution of the first-order wavelet index set $Q_{1}$, and averaging scale of the low-pass averaging filter $J$, to 8 each. 

In addition, we applied sample-wise min-max normalization to the input spectrogram and scattering coefficients and rescaled the intelligibility scores to 0-5 during training such that both the input features and output targets have similar scales. Finally, we evaluated our results using the mean square error (MSE), Pearson's correlation coefficient (PCC), and Spearman’s rank correlation coefficient (SRCC).

\subsection{Experimental results}

\subsubsection{Evaluation of intelligibility assessment}

First, we conducted an ablation study to test the performance of incorporating scattering coefficients. The results of In-MOSNet-S{\tiny II} and In-MOSNet-S{\tiny I} in Table~\ref{table:intell} show that scattering coefficients are more useful for intelligibility prediction than spectrograms. Also, the combination of scattering coefficients and spectrograms performs better than using only spectrograms or only scattering coefficients. 

Then, we tested the performance of incorporating quality scores into the intelligibility assessment. In Table~\ref{table:intell}, an \emph{In-} system has a similar model structure as an \emph{InQSS-} system but does not have a path to predict the quality scores. Comparing the performance of \emph{InQSS-} systems with \emph{In-} systems, the results show that the information regarding the quality scores can improve the performance of an intelligibility assessment. In addition, the ensemble model InQSS-MOSSSL achieves the best performance within the tested models.

Finally, we show the performance of STOI \cite{taal2011algorithm} and Google-ASR \cite{ref_google_asr} on the TMHINT-QI dataset. The Google-ASR results were calculated by computing the distance between the predicted and ground-truth sentences. The results in Table~\ref{table:intell} indicate the ASR results have a high correlation with the listening test results, whereas the STOI scores are less consistent with the listening test.


\begin{table}
\caption{Performance of different intelligibility assessment methods. Here, \emph{scat}, \emph{spec}, \emph{wav} are abbreviations of the scattering coefficients, spectrogram, and raw waveform, respectively.}
\centering
\begin{tabular}{|ccccc|} 
\hline Model / Method & Input& MSE   & PCC   & SRCC   \\ 
\hline\hline
In-MOSNet-S{\tiny I}& spec & 2.562 & 0.695 & 0.610  \\
In-MOSNet-S{\tiny II}& scat & 2.425 & 0.708 & 0.633  \\
In-MOSNet-S{\tiny III}& spec+scat & 2.393 & 0.714 & 0.642  \\
\rowcolor[rgb]{0.922,0.922,0.922} InQSS-MOSNet & spec+scat & 2.117 & 0.755 & 0.682  \\
\hline
In-SSL & wav & 2.571 & 0.749 & 0.645  \\ 
\rowcolor[rgb]{0.922,0.922,0.922} InQSS-SSL & wav & 2.552 & 0.754 & 0.664  \\
\hline
In-MOSSSL & 
\begin{tabular}[t]{@{}c@{}}wav\\spec+scat\end{tabular}
& 2.015 & 0.777 & 0.668  \\
\rowcolor[rgb]{0.922,0.922,0.922} InQSS-MOSSSL & \begin{tabular}[t]{@{}c@{}}wav\\spec+scat\end{tabular} & \textbf{2.017} & \textbf{0.791} & \textbf{0.700}  \\
\hline\hline
STOI \cite{taal2011algorithm} & - & 5.573 & 0.482 & 0.461  \\
Google-ASR & - & 7.305 & 0.710 & 0.679  \\
\hline\hline
\end{tabular}
\label{table:intell}
\end{table}

\subsubsection{Evaluation of quality assessment}

We tested the performance of incorporating intelligibility scores into the quality assessment. In Table~\ref{table:quality_results}, a \emph{Q-} system has a similar model structure as an \emph{InQSS-} system but does not have a path to predict the intelligibility scores. Compared the performance of \emph{InQSS-} systems with \emph{Q-} systems, the results indicate that the information regarding the intelligibility scores can improve the performance of the quality assessment. In addition, we notice that models trained on other datasets do not generalize well to the TMHINT-QI dataset. 

We then tested the model's performance on the DAPS dataset \cite{su2019perceptually}. 
The results listed in Table~\ref{table:quality_results} show that InQSS-MOSSSL performs better than Q-MOSSSL. However, the SSL model in
\cite{cooper2021generalization} obtains the best performance. The reason might be that the SSL model in \cite{cooper2021generalization} was trained on a much larger speech quality dataset than the TMHINT-QI, and therefore has a better generalizability than our model.

In addition, we find that MSE evaluation results are inconsistent with the PCC and SRCC results on out-of-domain datasets. One possible reason is that different datasets might have similar clean utterances but have diverse noisy utterances processed by different methods. Therefore, because the quality scores of noisy utterances in different datasets have discrepant scoring standards, models are less capable of predicting exact scores for out-of-domain data.



\begin{table}
\caption{Performance of different quality assessment models. In the model column, the star mark \emph{$*$} indicates whether the model has access to the training set of the dataset.}
\label{table:quality_results}
\centering
\begin{tabular}{|ccccc|} 
\hline
Model & Dataset & MSE & PCC& SRCC \\ 
\hline\hline
Q-MOSNet$^*$ & TMHINT-QI & 0.439 & 0.753 & 0.698 \\ 
\rowcolor[rgb]{0.922,0.922,0.922} InQSS-MOSNet$^*$ & TMHINT-QI & 0.422 & 0.763 & 0.715 \\
\hline
Q-SSL$^*$ & TMHINT-QI & 0.388 & 0.794 & 0.750 \\ 
\rowcolor[rgb]{0.922,0.922,0.922} InQSS-SSL$^*$ & TMHINT-QI & 0.365 & 0.800& 0.754 \\
\hline
\rowcolor[rgb]{0.922,0.922,0.922} InQSS-MOSSSL$^*$ & TMHINT-QI & \textbf{0.353} & \textbf{0.804} & \textbf{0.759}  \\ 
\hline
DNSMOS \cite{reddy2021dnsmos} & TMHINT-QI& 0.915 & 0.496& 0.311\\ 
NISQA \cite{mittag2021nisqa} & TMHINT-QI & 3.140 & 0.529 & 0.348 \\
SSL \cite{cooper2021generalization} & TMHINT-QI & 4.417 & 0.574 & 0.405 \\
\hline\hline
Q-MOSSSL & DAPS& 1.261& 0.617& 0.599  \\ 
\rowcolor[rgb]{0.922,0.922,0.922} InQSS-MOSSSL & DAPS & 1.100 & 0.639 & 0.639 \\
\hline
DNSMOS \cite{reddy2021dnsmos} & DAPS & 0.665 & 0.515& 0.510 \\
NISQA \cite{mittag2021nisqa} & DAPS & 0.663& 0.519 & 0.389 \\
SSL \cite{cooper2021generalization} & DAPS & \textbf{0.475} & \textbf{0.710} & \textbf{0.718}  \\
\hline\hline
\end{tabular}
\end{table}


\section{Conclusions}

In this study, we collected and analyzed the subjective and objective intelligibility and quality scores of clean, noisy, and enhanced utterances. Then, we released the dataset named TMHINT-QI. Moreover, we propose InQSS, a non-intrusive multi-task learning framework for intelligibility and quality assessment. The experimental results demonstrate that a multi-task learning network can improve the performance of a single task without increasing the model complexity. In addition, SSL-based models can achieve high performance on multi-task speech assessment and require less time to convergence than the training-from-scratch models. Finally, a simple ensemble approach, averaging the final predictions of two models, can effectively improve the results.


\bibliographystyle{IEEEtran}

\bibliography{ref}


\end{document}